\documentstyle[times,epsfig]{jaa}
\begin{document}
\title[Initiation and propagation of CMEs]{Initiation and propagation of
	coronal mass ejections}
\author[Chen]%
       {P. F. Chen\thanks{e-mail:chenpf@nju.edu.cn} \\ 
        Department of Astronomy, Nanjing University, Nanjing 210093,
        China}
\pubyear{2007}
\date{Received 2007 March; accepted 2007 May}
\maketitle
\label{firstpage}
\begin{abstract}
This paper reviews recent progress in the research on the initiation and
propagation of CMEs. In the initiation part, several trigger mechanisms are
discussed; In the propagation part, the observations and modelings of EIT 
waves/dimmings, as the EUV counterparts of CMEs, are described. 
\end{abstract}


\section{Introduction}
\label{intro}
Coronal mass ejections (CMEs) have been observed over 30 years. They keep
being an intriguing research topic, not only because they are now realized
to be the major driver for space weather disturbances, which are intimately
connected to human activities, but also because they themselves are full of
questions that have been provoking scientists to seek for answers. Stimulated
by the limited observations, theoretical researches are involved in the 
various phases of the eruptions from their birth to their pilgrimage in the
interplanetary (IP) space. First, we are still not quite sure what is the
progenitor of a CME. The rough picture is described as follows: magnetic 
field, which is generated at the tachocline layer, emerges throughout the
convection zone and the lower atmosphere into the tenuous corona. The coronal
field keeps adjusting to a more and more complex magnetic
structure in a quasi-steady way. After a threshold, the magnetic structure
can not sustain its equilibrium and begins to erupt. In this picture, it is
still an open question whether the pre-CME structure should always possess
a flux rope. Or, the so-called flux rope is actually an extreme case of the
ordinary magnetic arcade with a strong twist. The second issue is how the
progenitor is triggered to deviate from the equilibrium state. In this aspect,
the statistical investigations of the correlation between CME onsets and other
phenomena are of extreme significance. The third issue is how a CME is
accelerated. The related questions involve (1) whether magnetic reconnection
is a necessary condition, (2) how important the interaction between the ejecta
and the solar wind is, (3) the effect of prominence mass drainage, among 
others. The fourth issue is how the CME is related to the accompanied
phenomena, such as solar flares, Moreton waves, EIT waves and dimmings,
transient coronal holes, etc. The fifth issue is how the CME evolves to an
interplanetary CME (ICME) and how the CME properties affect the geomagnetic
activity.

In this review paper, we focus on two aspects of the theoretical researches on
CMEs, i.e., the initiation and propagation, which are presented Sections 2 and
3, respectively. We refer the readers to \nocite{forb00}Forbes (2000), 
\nocite{gopal03} Gopalswamy (2003), and Volume 123 of Space Sci. Rev. for
more detailed reviews. The chances and challenges of solar cycle 24 are
briefly prospected in Section 4.

\section{Initiation of CMEs}
\label{sec:ini}
Except some narrow CMEs, which may correspond to a jet (say, reconnection
jet) propagating along open field lines, most CMEs are regarded as an
erupting flux rope system, with a typical three-component structure in the
white-light coronagraph images, although sometimes one or two components
are absent possibly due to observational effect or the plasma has not yet
condensed to form a filament at the magnetic dips of the flux rope. The
eruption process can generally be described in the classical CSHKP framework:
a flux rope, which may or may not host a filament, becomes unstable or loses
its equilibrium, it then rises and pulls up the closed field lines straddling
over it, so as to form a current sheet beneath the flux rope. The reconnection
at the current sheet removes the constraint of the line-tied field lines, and
the flux rope is pushed to erupt by the upward reconnection jet. Therefore,
one important and unclear issue in this picture is how the flux system is
triggered.

\subsection{Emerging flux trigger mechanism}
\label{emer}
\begin{figure*}
\begin{center}
\psfig{file=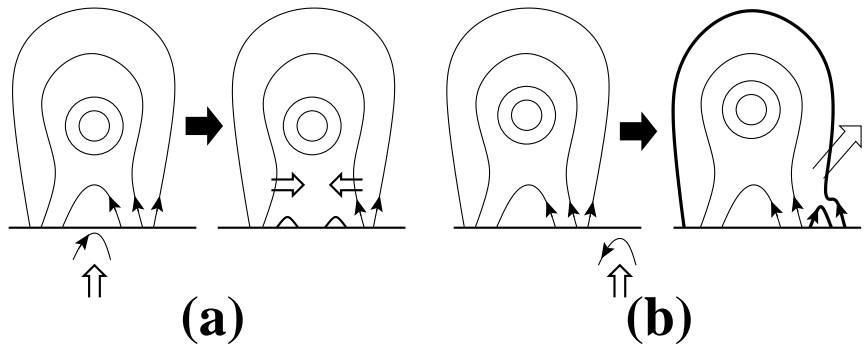, height= 4.5cm, angle=0}
\caption{Schematic diagram of the emerging flux trigger mechanism for CMEs.
(a) Emerging flux inside the filament channel cancels the pre-existing
loops, which results in the in-situ decrease of the magnetic pressure. Lateral
magnetized plasmas are driven convergently to form a current sheet; 
(b) Emerging flux outside the filament channel reconnects with the large
coronal loop, which results in the expansion of the loop. The underlying
flux rope then rises and a current sheet forms near the magnetic null point. }
\end{center}
\label{fig1}
\end{figure*}

Early in the 1970s, it was found that weak X-ray activities often precede
solar flares (Datlowe, Elcan, \& Hudson 1974), which were described as the
soft X-ray precursor of CMEs by Harrison et al. (1985). In an apparently
unrelated research, Feynman \& Martin (1995) found that many CMEs are strongly
associated with emerging flux that possesses polarity orientation favorable
for magnetic reconnection between the emerging flux and the pre-existing
coronal field either inside or outside the filament channel. \nocite{wang99}
Wang \& Sheeley (1999) confirmed the strong correlation between CMEs and
reconnection-favorable emerging flux, although it is noted that not all CMEs
are related to emerging flux. Motivated by such a correlation, we proposed an
emerging flux trigger mechanism for CMEs \nocite{chen00}(Chen \& Shibata
2000), as illustrated by Fig. \ref{fig1}: When the reconnection-favorable
emerging flux appears inside the filament channel, it cancels the small
magnetic loops near the polarity inversion line (PIL). Thereby, the magnetic
pressure decreases locally. Plasmas on both sides of the PIL, which are
initially in equilibrium, are driven to move convergently along with the
frozen-in anti-parallel magnetic field under the pressure gradient. As a
result, a current sheet forms above the PIL, and the flux rope is also 
triggered to move upward slightly. The
ensuing reconnection at the current sheet leads to the formation of a
cusp-shaped two-ribbon flare and the fast eruption of the flux rope system.
When the reconnection-favorable emerging flux appears outside the filament
channel (say, on the right side), it reconnects with the large-scale
magnetic arcades that cover the flux rope. The right leg of the arcade, which
is rooted very close to the PIL, is re-connected far from the PIL on the right
side of the emerging flux. The magnetic tension force along the curved field
line pulls the arcades to move upward, with the flux rope following
immediately. The rising flux rope pulls the overlying field lines up and a
current sheet forms near the null point below the flux rope. Similarly, the
magnetic reconnection at the current sheet leads to a two-ribbon flare and the
fast eruption of a CME. 

In this model, the onset of the CME is triggered by the localized reconnection
between emerging flux and the pre-existing coronal field. Such a reconnection
produces X-ray jets (and H$\alpha$ surges if chromosphere is considered as
done in \nocite{yoko95} Yokoyama \& Shibata 1995), which correspond to the
soft X-ray precursor well before the main flare as mentioned by 
\nocite{harr85}Harrison et al. (1985). The numerical results also show that the impulsive phase of
the main flare coincides with the acceleration phase of the CME, and after
the flare peak, the CME moves with an almost constant velocity. \nocite{jing04}
Jing et al. (2004) found that about 68\% of disk CMEs are associated with
emerging flux. So, the onset of quite a large part of the CMEs can be explained
by our model. The simulation results in this model were also found to be
consistent with various observations (e.g., \nocite{zhan01}Zhang et al. 2001). In particular, 
\nocite{ster05} Sterling \& Moore (2005) analyzed a CME event,
in which they found that the height profile of the filament is very similar to
that in our paper.   A parameter survey of this model was conducted
by \nocite{xu05}Xu, Chen, \& Fang (2005); its image synthesis was composed
by \nocite{shio05} Shiota et al. (2005) in order to compare with Yohkoh/SXT
images. Such a model was recently extended to the spherical coordinators
\nocite{dub05}(Dubey, Holst, \& Poedts 2006).

\subsection{Other trigger mechanisms}
Observations have shown various kinds of evolving magnetic structures, for
example, converging motion of the filament channel (note that the apparent
converging motion may be a result of the diffusion of magnetic polarities),
shear motion, twist motion, decay of the active region, etc. Accordingly,
trigger mechanisms based on these changes have also been proposed. 
\nocite{van89}van Ballegooijen and Martens (1989) proposed that the converging
motion of magnetic arcades, by which a filament may be formed, can also lead
to the destabilization of the filament; \nocite{miki88}Miki\'c, Barnes, \&
Schnack (1988) found that after large enough shear, the closed magnetic
arcades would asymptotically approach the open field, while a resistive
instability can result in the eruption. \nocite{kusa04}Kusano et al. (2004),
however, found that reversed magnetic shear could also trigger the eruption.
Both analytical and numerical
simulations indicate that there may exist catastrophic behavior in the flux
rope motions as the footpoints of the magnetic arcades converge or shear
\nocite{forb95,hu01}(Forbes \& Priest 1995; Hu \& Jiang 2001). The analytical
investigation by \nocite{isen93}Isenberg, Forbes, \& Demoulin (1993)
illustrates that the gradual decay of the background magnetic field would also
cause the flux rope to lose equilibrium catastrophically. In all of these
three cases and our emerging flux model, the essence is that the evolving
magnetic structure either increases the magnetic pressure below the flux rope
or decreases the magnetic tension force above the flux rope, thereby the flux
rope cannot sustain its equilibrium.

\nocite{chen97}Chen et al. (1997) and \nocite{kral01}Krall et al. (2001)
proposed that the injection of poloidal magnetic flux into the flux rope
would cause the flux rope to erupt. Physically this process is similar to
the kink instability model as put forward by \nocite{hood81}Hood \& Priest
(1981) and simulated by \nocite{toro05}. The model has been compared with
observations in many cases. A modified version of the kink model, i.e., the
rupture mechanism, was proposed by \nocite{stur01}Sturrock et al. (2001) and
simulated by \nocite{fan05}Fan (2005), where part of the flux rope penetrates
the overlying magnetic field and erupts into the IP space.

In order to circumvent the Aly's constraint, \nocite{anto89}Antiochos, DeVore,
\& Klimchuk (1999) proposed a magnetic breakout model, i.e., only the sheared
part of the closed field lines near the PIL is opened during the CME. The
essence of this model is that the overlying background magnetic field
reconnects with the sheared arcade at the magnetic null point above the
latter, by which the constraint over the sheared arcade is removed 
gradually like an onion-peeling process. If such a reconnection above the
arcade exists during the onset of the CME, it is expected to see soft X-ray
bright loops on both sides of the sheared arcade and inverse type III radio
bursts that are produced by the reconnection-accelerated electrons.

There are some other less-recognized trigger mechanisms for CMEs. Filament
mass drainage, by which the filament obtains a buoyancy force, may play a
role in triggering the onset of a CME (Low 2001), which was identified
in one event recently (Zhou et al. 2006). Moreton and EIT waves, which are
generated by a remote CME, often trigger a filament to oscillate, and
erupt sometimes (Ballester 2006), which deserves further investigations.

\section{Propagation of CMEs}
\label{pro}
As mentioned in Section \ref{intro}, there are many interesting topics
related to the propagation of CMEs. Here, we just mention the EUV counterparts
of propagating CMEs, i.e., EIT waves/dimmings.

\subsection{EIT wave/dimming observations}
EIT waves were originally observed by the EIT telescope on board the SOHO
satellite as propagating wave-like fronts, with an emission enhancement
ranging from 25\% to less than 14\%, which is followed immediately by
expanding EIT dimmings (Thompson et al. 1998).
Therefore, EIT waves and dimmings are symbiotic phenomena. One typical feature
is that the bright fronts propagate only in the quiet regions, avoiding any
active region. Therefore, when the large-scale magnetic configuration is
simple, for instance, with only one active region on the visible disk, the EIT
wave fronts are almost circular; however, when there are other active regions
surrounding the source region of the eruption, the EIT waves appear in
patches, managing their ways outward separately in the quiet regions.

\nocite{will99}Wills-Davey \& Thompson (1999) found that EIT waves can be
observed in both 195 {\AA} (with the formation temperature $T\sim$ 1.4 MK) and
171  {\AA} (with the formation temperature $T\sim$ 1 MK), with more detailed
structures in the 171  {\AA} images. Later, \nocite{zhuk04}Zhukov \& Auchere
(2004) also identified EIT waves in 284 {\AA} (with the formation temperature
$T\sim$ 1.9 MK). Since the brightenings are observed at very different
temperatures, it is concluded that they are mainly due to the density
enhancement, although \nocite{will99}Wills-Davey \& Thompson (1999) and
\nocite{chen05}Chen \& Fang (2005) pointed out that temperature effect is not
negligible. Weak dimmings are also reported at 304 {\AA} (\nocite{cher03}
Chertok \& Grechnev 2003). However, it is not sufficient to say they 
have imprints in the chromosphere since a coronal line Si XI 303.32 {\AA} and
a transition region line He II 303.78 {\AA} are blended at the EIT 304 bandpass.
It is generally believed that EIT waves are a phenomenon propagating in the
corona. Based on the observational results in Thompson et al. (2000) and
Harrison et al. (2003), Chen \& Fang (2005) proposed that the ``EIT waves"
map the footprints of the CME leading edge, and the dimming region maps the
the bottom of the CME cavity. Therefore, EIT waves/dimmings are actually the
EUV counterparts of CMEs.

\subsection{Debates on EIT wave mechanism}

Early in the 1960s, it was discovered in the H$\alpha$ line wing  that
arc-shaped chromospheric perturbations propagate away from some big flares
(Moreton \& Ramsey 1960), which were later called Moreton waves. Such a wave,
with
a surprisingly large velocity on the order of 1000 km s$^{-1}$, was later explained
by Uchida (1968) as a fast-mode wave in the corona, which sweeps the
chromosphere as it propagates. Since then, it was expected to detect such
a wave in the corona. It was not successful except one event observed by
OSO 7 satellite (Neupert 1989). The discovery of EIT waves by SOHO/EIT, then,
sparked a lot of interests, as well as controversies.  
It was very natural to consider the EIT waves as the ever-missing coronal
counterparts of Moreton waves (or coronal Moreton waves for short), i.e.,
they are coronal fast-mode waves.
After extrapolating the coronal magnetic field based on a potential field
model, Wang (2000) and Wu et al. (2001) claimed that the propagating 
fast-mode waves in the corona can match the observed EIT wave fronts.
However, it is very difficult for the fast-mode wave model to explain the
typical features of EIT waves: (1) The EIT wave speeds are 3 or more
times smaller than Moreton waves (Klassen et al. 2000); (2) Delann\'ee \& 
Aulanier (1999) found that EIT waves stop at the magnetic separatrix, which
led them to speculate that EIT waves could be associated with magnetic
rearrangement; (3) The EIT velocities are not correlated with the speeds of
the type II radio bursts, the latter of which are believed to be the radio
signature of the coronal fast-mode shock waves; (4) EIT wave speeds can be
as low as 50 km s$^{-1}$ (Thompson \& Myers 2007), which is even below the
sound speed in the corona. However, fast-mode wave speed should always be
larger than the sound speed.

In order to reconcile all these discrepancies, \nocite{chen02}Chen et al.
(2002) and \nocite{chen05}Chen, Fang, \& Shibata (2005) predict that there
should exist two EUV waves associated with a CME event, i.e., the coronal
Moreton wave and the EIT wave, which was later confirmed by \nocite{harr03}
Harra \& Sterling (2003). In our model, the coronal Moreton waves correspond
to the piston-driven shock over the CME rather than the blast wave from the
pressure pulse in the flare, and EIT waves are generated by successive
opening (or stretching) of closed field lines, which is pushed by the erupting
flux rope.  Each field line is pushed to expand at its top, and the
deformation is transferred down to the footpoints of the field line. Whenever
the leg of a field line expands, the plasma outside the field line is
compressed to form an EIT wave front, while the plasma inside is evacuated,
resulting in EIT dimmings. Therefore, the model can explain both EIT waves and
dimmings.
The numerical results reproduce many characteristics that are obtained from
observations: (1) EIT waves propagate with a velocity $\sim$3 times smaller
than the coronal Moreton waves; (2) EIT waves stop at the magnetic separatrix
between the source active region and another active region; (3) the EIT wave
speed is anti-correlated with the speed of type II radio bursts.

\subsection{Significance of EIT wave/dimming observations}

(1) EIT waves/dimmings are the disk signatures of CMEs: \nocite{bies02}
Biesecker et al. (2002) found that whenever there is an EIT wave, there should be a CME in the
coronagraph images, although the contrary is not true. As mentioned above,
EIT waves/dimmings map the CME leading edge/cavity, and they are the disk
signatures of the CMEs. Therefore, routine observations of EIT
waves/dimmings will be an efficient way to monitor CMEs, especially
those directed toward our Earth;

(2) EIT dimmings provide an estimate of the mass supply for CMEs: CMEs are the
major driver for space weather disturbances such as geomagnetic storms. 
Their mass, as well as their velocity and the magnetic field, is an important
factor that may influence their geomagnetic effect. Hence, the estimate of
their mass in the early phase of the eruption is crucial for space weather
forecast. Harrison \& Lyons (2000) proposed that EIT dimmings, which are due
to the plasma evacuation as in our model, can be used to estimate the CME mass;

(3) Large-scale coronal magnetic field can be inferred: Various efforts have
been put into the measurement of the coronal magnetic field, such as the
radio diagnosis, near infrared Zeeman effect measurements, and so on. Before
these methods become practical, EIT waves/dimmings can provide an efficient
way to diagnose the coronal magnetic field. As discussed in Chen et al. (2002)
and Chen, Fang, \& Shibata (2005), EIT waves/dimmings are produced by the
opening of the closed field lines covering the erupting flux rope. This means 
that coronal field lines should be self-closed within the dimming regions.
With sufficiently high cadence of the EIT wave observations, their velocity
pattern can even be used to derive the coronal magnetic field.

\section{Prospects for the Colar Cycle 24}

It is seen from the above review that our understanding of the CME
initiation and propagation strongly relies on observations. For the CME
initiation, it will be vital to detect the progenitor of the CME and its
early evolution in order to distinguish between various trigger models.
In this sense, UV coronagraph observations would be invaluable to trace the
early evolution of any ongoing eruption; on the other hand, sub-surface
detections of the magnetic field and motions based on the local seismology
would also be helpful. For the EIT waves/dimmings, we believe that the 
ongoing STEREO/SECCHI observations with a high cadence would gradually
uncover the veil over the spectacular phenomenon, which can then be used as
the proxy for the coronal magnetic field diagnosis.

\section*{Acknowledgements}
The author thanks the referee for the comments. The research is supported by
the Chinese foundations NCET-04-0445, FANEDD
(200226), 2006CB806302, NSFC (10221001, 10333040, 10403003, and 106100099).

\newpage
\label{lastpage}
\end{document}